\documentclass[10pt,journal]{IEEEtran}

\usepackage{comment}
\usepackage{pgfplots}
\usepackage{pgfplotstable}
\pgfplotsset{compat=1.13}
\usepackage{tikz}
\usetikzlibrary{pgfplots.groupplots}
\usetikzlibrary{pgfplots.statistics}
\usetikzlibrary{calc,tikzmark}
\usepackage{enumitem}
\newlist{rqs}{enumerate}{1}
\setlist[rqs,1]{label={\bfseries RQ\arabic*},
itemindent=0.5em}

\usepackage{fancyvrb}
\usepackage[ruled,vlined,linesnumbered]{algorithm2e}
\usepackage{xspace}
\usepackage{listings}
\usepackage{balance}
\usepackage{array}
\usepackage[normalem]{ulem}
\usepackage{booktabs}
\usepackage{subcaption}
\usepackage{caption}
\usepackage{framed}
\usepackage{algorithmic}
\usepackage{graphicx}
\usepackage{textcomp}
\usepackage{xcolor}
\usepackage{tcolorbox}
\usepackage{soul}
\usepackage{multirow}
\usepackage{pifont}
\usepackage{tcolorbox}
\usepackage{makecell}
\usepackage{bbm}
\usepackage{MnSymbol}
\usepackage{amsfonts}
\usepackage{threeparttable}
\usepackage{adjustbox}
\usepackage{boldline}

\usepackage[sort,numbers]{natbib}
\definecolor{LinkGreen}{rgb}{0.0, 0.5, 0.0}
\definecolor{CiteRed}{rgb}{0.8, 0.0, 0.0}
\definecolor{LinkMaron}{rgb}{0.5, 0.0, 0.0}
\usepackage[colorlinks, citecolor=CiteRed, linkcolor=LinkGreen, urlcolor=LinkMaron]{hyperref}

\makeatletter
\let\old@lstKV@SwitchCases\lstKV@SwitchCases
\def\lstKV@SwitchCases#1#2#3{}
\makeatother
\usepackage{lstlinebgrd}
\makeatletter
\let\lstKV@SwitchCases\old@lstKV@SwitchCases

\lst@Key{numbers}{none}{
    \def\lst@PlaceNumber{\lst@linebgrd}
    \lstKV@SwitchCases{#1}
    {none:\\%
     left:\def\lst@PlaceNumber{\llap{\normalfont
                \lst@numberstyle{\thelstnumber}\kern\lst@numbersep}\lst@linebgrd}\\%
     right:\def\lst@PlaceNumber{\rlap{\normalfont
                \kern\linewidth \kern\lst@numbersep
                \lst@numberstyle{\thelstnumber}}\lst@linebgrd}
    }{\PackageError{Listings}{Numbers #1 unknown}\@ehc}}
\makeatother

\usepackage[capitalize, nameinlink]{cleveref}
\newcommand\defbf[1]{\textcolor{black}{#1}}
\crefdefaultlabelformat{#2\textbf{#1}#3}
\crefname{section}{\defbf{Section}}{\defbf{Sections}}
\crefname{table}{\defbf{Table}}{\defbf{Tables}}
\crefname{figure}{\defbf{Figure}}{\defbf{Figures}}
\crefname{subfigure}{\defbf{Figure}}{\defbf{Figures}}
\crefname{definition}{Definition}{Definitions}
\crefname{equation}{Equation}{Equations}
\crefname{example}{Ex.}{Examples}
\crefname{algorithm}{Algorithm}{Algorithms}
\crefname{line}{Line}{Lines}

\ExplSyntaxOn
\newcommand{\calculate}[1]{\fp_eval:n { #1 }}
\ExplSyntaxOff
\newcommand{\ratio}[2]{$\calculate{round({#1} /{#2} *100, 2)}\%$}

\newcommand{\csrewite}[1]{{\color{black}{#1}}}

\newcommand{\cmark}{\ding{51}}
\newcommand{\xmark}{\ding{55}}
\newcommand{\gxmark}{{\color{lightgray}\xmark}}

\newcommand{\boldbullet}[1]{$\bullet$ The \textbf{#1}}

\newcommand{\scalelens}[0]{\textsc{ScaleLens}\xspace}
\newcommand{\scaleview}[0]{\textsc{ScaleView}\xspace}
\newcommand{\scalepick}[0]{\textsc{ScalePick}\xspace}
\newcommand{\scalecheck}[0]{\textsc{ScaleCheck}\xspace}
\newcommand{\sfind}[0]{\textsc{SFind}\xspace}
\newcommand{\stest}[0]{\textsc{STest}\xspace}

\newcommand{\codeql}[0]{\textsc{CodeQL}\xspace}

\newcommand{\eg}{{e.g.,}\xspace}
\newcommand{\ie}{{i.e.,}\xspace}

\newcommand{\etc}{{etc.}\xspace}

\definecolor{javared}{rgb}{0.6,0,0}
\definecolor{javagreen}{rgb}{0.25,0.5,0.35}
\definecolor{javapurple}{rgb}{0.5,0,0.35}
\definecolor{javadocblue}{rgb}{0.25,0.35,0.75}

\newcommand{\listingsttfamily}{\fontfamily{SourceCodePro-TLF}}

\lstdefinestyle{ShellCode}{
  language=sh,
  columns=fullflexible,
  basicstyle=\ttfamily\scriptsize,
  keywordstyle=\color{javapurple}\bfseries,
  commentstyle=\listingsttfamily\color{javagreen}\itshape,
  numbers=left,
  numbersep=0pt,
  keywords={for,do,done},
  rulecolor=\color{gray},
  frame=r,
  framesep=-40pt,
}
\lstdefinestyle{JavaCode}{
  language=Java,
  columns=fullflexible,
  basicstyle=\ttfamily\scriptsize,
  keywordstyle=\color{javapurple}\bfseries,
  commentstyle=\listingsttfamily\color{javagreen}\itshape,
  numbers=left,
  numbersep=-12pt,
  rulecolor=\color{gray},
  frame=r,
  framesep=-2pt,
  xleftmargin=-16pt,
  tabsize=4,
  captionpos=b,
  showlines=true,
}

\lstdefinestyle{JavaCodeNoLine}{
  language=Java,
  columns=fullflexible,
  basicstyle=\ttfamily\scriptsize,
  keywordstyle=\color{javapurple}\bfseries,
  commentstyle=\listingsttfamily\color{javagreen}\itshape,
  rulecolor=\color{gray},
  frame=r,
  framesep=-2pt,
  xleftmargin=-16pt,
  tabsize=4,
  captionpos=b,
  showlines=true,
  escapechar=~,
  breaklines=true,
  postbreak=\mbox{\textcolor{red}{$\hookrightarrow$}\space},
}

\lstdefinestyle{CSV}{
  language=Java,
  columns=fullflexible,
  basicstyle=\ttfamily\scriptsize,
  rulecolor=\color{gray},
  escapechar=~,
}

\lstdefinestyle{JavaDiff}{
  language=Java,
  columns=fullflexible,
  basicstyle=\listingsttfamily\scriptsize,
  keywordstyle=\color{javapurple}\bfseries,
  commentstyle=\listingsttfamily\color{javagreen}\itshape,
  xleftmargin=0pt,
  captionpos=b,
  escapechar=~,
  breaklines=true,
  postbreak=\mbox{\textcolor{red}{$\hookrightarrow$}\space},
}

\def \xs {\xspace}

\def \io {IO\xs}
\def \os {OS\xs}

\def \oom {OOM\xs}

\def \dcfs {DCFs\xs}
\def \dcf {DCF\xs}

\def \sysca {Cassandra\xs}
\def \syshd {HDFS\xs}
\def \sysha {Hadoop\xs}
\def \sysyr {Yarn\xs}
\def \sysmr {MapReduce\xs}
\def \sysst {Storm\xs}
\def \syssp {Spark\xs}
\def \sysig {Ignite\xs}
\def \syskf {Kafka\xs}
\def \syshb {HBase\xs}
\def \syszk {ZooKeeper\xs}
\def \numrepos {10\xs}
\def \totalbugs {199K\xs}

\def \numbugs {444\xs}

\def \numload {154\xs}
\def \numfail {16\xs}
\def \numdata {176\xs}
\def \numclus {98\xs}

\def \numcompute {196\xs}
\def \percompute {44.1\%\xs}
\def \numcomputecross {63\xs}
\def \percomputecross {32.1\%\xs}
\def \numcomputesync {76\xs}
\def \percomputesync {38.8\%\xs}
\def \numcomputeapp {57\xs}
\def \percomputeapp {29.1\%\xs}
\def \numunb {103\xs}
\def \perunb {23.2\%\xs}
\def \numunbcollection {44\xs}
\def \perunbcollection {42.7\%\xs}
\def \numunballoc {38\xs}
\def \perunballoc {36.9\%\xs}
\def \numunbos {21\xs}
\def \perunbos {20.4\%\xs}
\def \numbloat {32\xs}
\def \perbloat {7.2\%\xs}
\def \numbloatwaste {20\xs}
\def \perbloatwaste {62.5\%\xs}
\def \numbloatopt {12\xs}
\def \perbloatopt {37.5\%\xs}
\def \numlogic {113\xs}
\def \perlogic {25.5\%\xs}
\def \numlogicleak {64\xs}
\def \perlogicleak {56.6\%\xs}
\def \numlogicrace {17\xs}
\def \perlogicrace {15.1\%\xs}
\def \numlogiccorner {32\xs}
\def \perlogiccorner {28.3\%\xs}

\newcommand{\yr}[1]{\jirabib{YARN}{YR-}{#1}}
\newcommand{\ca}[1]{\jirabib{CASSANDRA}{CA-}{#1}}

\newcommand{\ha}[1]{\jirabib{HADOOP}{HA-}{#1}}
\newcommand{\hd}[1]{\jirabib{HDFS}{HD-}{#1}}
\renewcommand{\st}[1]{\jirabib{STORM}{ST-}{#1}}
\newcommand{\ig}[1]{\jirabib{IGNITE}{IG-}{#1}}
\newcommand{\kf}[1]{\jirabib{KAFKA}{KF-}{#1}}
\newcommand{\jira}[3]{\href{http://issues.apache.org/jira/browse/#1-#3}{#2#3}}
\newcommand{\jirabib}[3]{\href{http://issues.apache.org/jira/browse/#1-#3}{#2#3}~\citep{#1-#3}}

\newcommand{\yrnc}[1]{\jira{YARN}{YR-}{#1}}
\newcommand{\canc}[1]{\jira{CASSANDRA}{CA-}{#1}}

\newcommand{\hanc}[1]{\jira{HADOOP}{HA-}{#1}}
\newcommand{\hdnc}[1]{\jira{HDFS}{HD-}{#1}}

\newcommand{\ignc}[1]{\jira{IGNITE}{IG-}{#1}}
\newcommand{\kfnc}[1]{\jira{KAFKA}{KF-}{#1}}

\def \benchtotal {55\xs}

\begin{document}

\title{{Understanding and Detecting Scalability Faults in Large-Scale Distributed Systems}}

\author{Hao-Nan Zhu$^{*}$, Goodness Ayinmode$^{*}$, Cesar A. Stuardo, Haryadi S. Gunawi, and Cindy Rubio-Gonz\'{a}lez
\thanks{$^{*}$Contributed equally. H.-N. Zhu, G. Ayinmode, and C. Rubio-Gonz\'{a}lez are with the University of California, Davis, United States (e-mail: hnzhu@ucdavis.edu; goayinmode@ucdavis.edu; crubio@ucdavis.edu). C. A. Stuardo and H. S. Gunawi are with the University of Chicago, United States (e-mail: cesar.stuardo@bytedance.com; haryadi@cs.uchicago.edu).}
}

\maketitle

\begin{abstract}
  Scalable distributed systems form the backbone of modern
  computing infrastructure. However, as scale grows, system complexity
  may lead to scalability faults. Scalability faults are challenging
  to uncover and diagnose, as they are often latent and only
  manifest at large-scale deployment.
  \csrewite{In this paper, we present
    the first comprehensive study on scalability faults and propose
    an approach for their detection. First, we systematically
    investigate \numbugs scalability issue reports from \numrepos
    large-scale distributed systems to understand the common
    anti-patterns and root causes of scalability faults. We found that
    the majority of these faults are caused by the synergy
    between dimensional code fragments and anti-patterns
    associated with them. Second, based on our findings, we design
    and implement \scalelens, a novel approach to detect scalability faults.
    \scalelens combines dynamic and static analyses to pinpoint dimensional
    code fragments and match them with anti-patterns.
    Our evaluation shows that \scalelens detects 4.2\texttimes\xspace
    more dimensional code fragments associated with known scalability faults
    compared to the baseline. On the latest stable
    versions of \sysca, \syshd, and \sysig, \scalelens detects \calculate{129+68+137} dimensional code
    fragments with confirmed problematic behavior.}
\end{abstract}

\begin{IEEEkeywords}
  Scalability, distributed systems, fault detection, program analysis.
\end{IEEEkeywords}

\section{Introduction}
\label{sec:intro}

\IEEEPARstart{G}{iven} the constraints of Moore's Law and Dennard scaling, coupled with
the escalating demand for computing power, the last decade has seen
unprecedented deployment scales: over 300,000 \sysca nodes are hosting
over 100 petabytes of data at Apple, tens of thousands of
\sysmr/\syssp jobs are running on the \sysha clusters of
$\mathbbmss{X}$ (Twitter), and trillions of \syskf messages are
processed per day at
LinkedIn~\citep{twitterengineering.web,applecassandra.web,linkedinkafka.web}.
With highly-scalable distributed systems~\citep{apacheprojects.web}, one can
surpass the limitations of a single machine in meeting the increasing demand in
computation and storage.
To accommodate growth, distributed systems employ two fundamental scaling
strategies: \textit{scale-out} (horizontal scaling), which adds more nodes to
the cluster, and \textit{scale-up} (vertical scaling), which increases resources
per node.
While these systems are inherently designed to scale, their complexity makes
them prone to \textit{scalability faults}.

Scalability faults are defined as faults that manifest
only when one or more system aspects (\eg number of nodes, data volume, concurrent requests)
increase beyond certain thresholds,
but remain latent at smaller scales~\citep{DBLP:conf/hotos/Leesatapornwongsa17}.
Scalability faults are orthogonal to traditional fault categories such as
functional, performance, security, or concurrency faults.
A scalability fault may exhibit symptoms similar to these categories
(\eg performance degradation, resource exhaustion, or race conditions),
but is fundamentally characterized by its \textit{scale-dependent manifestation}:
the fault remains hidden during small-scale testing and only emerges under
large-scale conditions.
Revealing scalability faults requires large-scale deployment, which is
difficult and expensive to achieve during the development and testing
phases. As a result, scalability faults are often discovered in
production environments~\cite{amazon11201.web,amazon12721.web,amazon5467D2.web,circleci.web,googleincidentX8SNk.web,googleincidentmREMLw.web},
leading to resource shortage, performance
degradation, or even system crashes. 

\begin{figure}[!t]
    \centering
    \includegraphics[width=0.95\columnwidth]{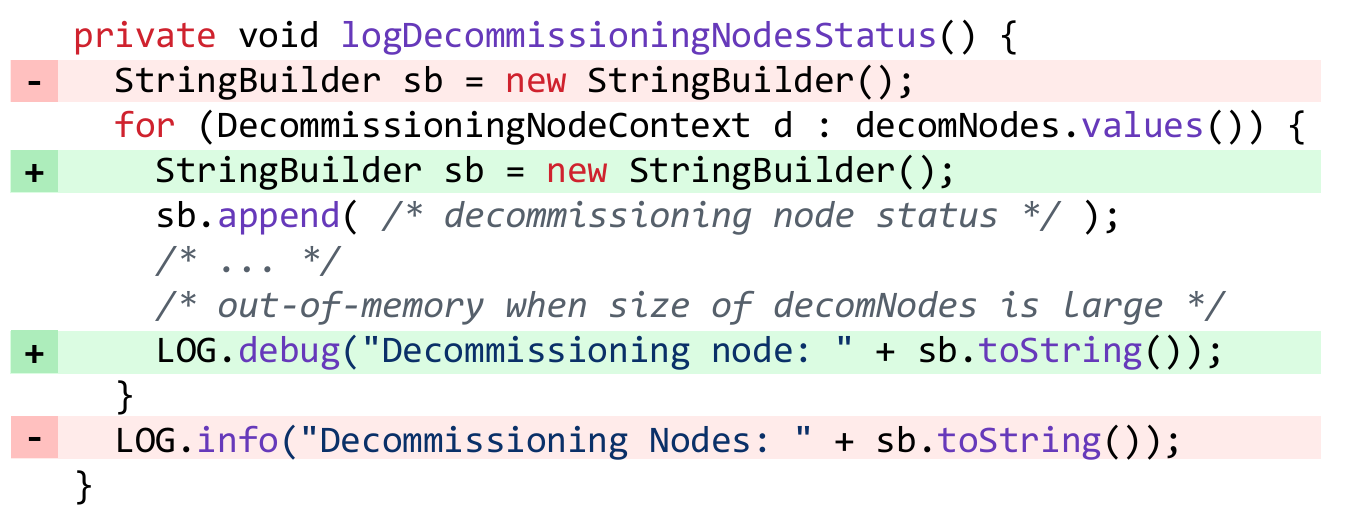}
    \caption{\yr{6188} : \textit{``... and yes, it does throw an OOM exception in case of large clusters"}}
    \label{fig-yr-6188}
\end{figure}

\cref{fig-yr-6188} shows an
example of a scalability fault and its fix in \sysha \sysyr. In
\yr{6188}, developers notice that the unbounded design of a data
structure and a large number (\ie $\geq 2000$) of decommissioning
nodes lead to a system crash due to an out-of-memory (\oom) error. The
root cause is the combination of (1) the \texttt{for} loop whose
number of iterations grows with the number of decommissioning nodes,
and (2) an unbounded data structure \texttt{StringBuilder} whose size
grows in every iteration of the loop. If the number of decommissioning
nodes is large, the \texttt{StringBuilder} object will exhaust memory
and trigger an \oom error. The fix is to move the
\texttt{StringBuilder} object inside the \texttt{for} loop to leave it
out of scope after each iteration, thus allowing the garbage collector
to free up memory.
This example illustrates the scale-dependent nature of scalability faults:
the code functions correctly with a small number of decommissioning nodes
but fails catastrophically when that number exceeds a threshold.

\definecolor{lightgrey}{rgb}{0.9, 0.89, 0.89}
\definecolor{diffpink}{rgb}{0.99, 0.71, 0.71}
\definecolor{diffgreen}{rgb}{0.71, 0.99, 0.71}

\begin{table*}[!t]

    \centering
    \caption{Breakdown of Faults in the Empirical Study.}
    \scriptsize
    \begin{tabular}{ll|rrrrrrrrrr|r}
        \toprule
        Dimension & Examples                               & \sysca & \sysha & \syshb & \syshd & \sysig & \syskf & \sysmr & \syssp & \sysst & \sysyr & \textit{Total} \\
        \midrule
        Load       & \# requests, RPCs, jobs                & 9  & 15 & 13 & 36 & 13 & 9  & 22 & 15 & 11 & 11 & \numload       \\
        Data       & \# tables, files, dirs, or their sizes & 29 & 10 & 12 & 49 & 14 & 21 & 7  & 27 & 1  & 6  & \numdata       \\
        Cluster    & \# peers, datanodes, namenodes         & 29 & 5  & 9  & 17 & 17 & 14 & 2  & 1  & 2  & 2  & \numclus       \\
        Failure    & \# node failures                       & 0  & 2  & 2  & 3  & 1  & 1  & 4  & 1  & 1  & 1  & \numfail       \\
        \midrule
                   & \multicolumn{1}{r|}{\textit{Total}}    & 67 & 32 & 36 & 105& 45 & 45 & 35 & 44 & 15 & 20 & \textbf{\numbugs}   \\
        \bottomrule
    \end{tabular}
    \label{tab-dim-systems}
\end{table*}

Unfortunately, scalability faults are not well understood. Previous
work~\citep{DBLP:conf/hotos/Leesatapornwongsa17} has revealed their
existence and provided a high-level overview of their potential
impacts based on a small number of observations. However, many
questions remain unanswered: Which aspects of a system cause
scalability faults when scaled? Are there common root causes and/or
code anti-patterns that induce scalability faults? Is it possible to
systematically detect scalability faults without large-scale
deployment? Each of these questions implies significant challenges, as
scalability faults are often difficult to reproduce and diagnose
unless the system is deployed at large scale.

To address the above questions, we conduct a comprehensive empirical study
encompassing \numbugs scalability faults from \numrepos large-scale
distributed systems.
For each scalability fault, we manually inspect the issue report to identify (1)
which system aspects are being scaled, and (2) the root causes and/or common
anti-patterns that may induce scalability faults.  

From (1), we report the aspects of
the system that are subject to change in scale and significantly impact the
system's scalability with their growth.  We refer to such aspects as
\textbf{scalable system dimensions}~\cite{DBLP:conf/icse/DubocRW06,
DBLP:conf/sigsoft/DubocRW07}, which we categorize into four kinds:
(i)~\textit{load}, the volume of work the system processes (\eg \#requests, RPCs, jobs);
(ii)~\textit{data}, the volume of items the system stores (\eg \#tables, files, directories, or their sizes);
(iii)~\textit{cluster}, the size of the deployment (\eg \#peers, datanodes, namenodes); and
(iv)~\textit{failure}, the number of concurrent component failures the system must tolerate (\eg simultaneously failing nodes during recovery).
\cref{tab-dim-systems} lists examples of each scalable system dimension and
their corresponding number of faults. 

From (2), we identify four
root-cause categories of scalability faults:
(i)~\textit{compute}, execution-time overhead tied to scale;
(ii)~\textit{unbound}, resource-consumption overhead tied to scale;
(iii)~\textit{bloat}, scalability-limiting data structure designs; and
(iv)~\textit{logic}, logic flaws that surface only at large scale.
Within these categories, we further identify 11 \textbf{anti-patterns},
\ie concrete code or design forms that repeatedly induce scalability faults, with two or three
anti-patterns per category, described individually in
\cref{sec-empirical-fault-patterns}.
Across these anti-patterns, we observe that
\textbf{dimensional code fragments (\dcfs)} play a fundamental role.
A \dcf is an iterative code fragment whose number of iterations is
correlated with one or more scalable system dimensions. For example,
the \texttt{for} loop in \cref{fig-yr-6188} is a \dcf whose iteration
count grows with the number of decommissioned nodes.
\dcfs are common and often necessary in scalable systems, and are neither
faults nor anti-patterns by themselves; they become fault-relevant only when
associated with certain anti-patterns
(\eg the unbounded \texttt{StringBuilder} in \cref{fig-yr-6188}).
In our study, \dcfs underlie \ratio{299}{444} of scalability faults in the
\textit{compute} and \textit{unbound} categories; the remaining
\ratio{145}{444} in the \textit{bloat} and \textit{logic} categories involve
other mechanisms.

Then, by formalizing the insights from our empirical study, we design and
implement \scalelens, a \dcf-centered approach to detect scalability faults
and identify their root causes while utilizing a single machine.
\scalelens consists of two main components: \scaleview
and \scalepick, which perform dynamic and static analyses,
respectively.
\scaleview uses runtime instrumentation together with
scaling workloads to collect execution traces. These traces
are then automatically analyzed to
determine correlations between scalable system
dimensions and code fragments to identify
\dcfs and categorize their computational complexity.
\scalepick performs call-graph analysis
on the DCFs to identify anti-patterns and pinpoint fragments that lead to scalability faults.

Our evaluation of \scalelens is two-fold.
First, we evaluate \scalelens on \benchtotal previously known real-world scalability
faults from \sysca, \syshd, and \sysig, which are widely-used distributed
systems. Our evaluation shows that \scalelens is successful at fully
detecting 36 (and partially detecting 2 more) out of \benchtotal scalability
faults, outperforming the baseline by 4.2\texttimes\xspace.
Second, we evaluate the capability of \scalelens to detect previously unknown scalability faults.
Specifically, we use \scalelens to analyze the latest stable
versions of \sysca, \syshd, and \sysig, in which it detects
a total of \calculate{129+68+137} \dcfs with associated anti-patterns.
After manual inspection, we find that all cases constitute problematic
behaviors. We are in the process of
reporting the new faults; so far 5 of reported \dcfs have been confirmed by developers, and 4 more are under investigation.

In summary, this paper makes the following contributions:

\begin{itemize}
    \item We conduct the largest-to-date study on scalability faults,
          from which we derive the existence of \dcfs and
          conclude common scalability anti-patterns
          (\cref{sec-empirical-fault-patterns}).
    \item \csrewite{We design and implement \scalelens, a novel
              approach that combines dynamic and static
              analyses to detect scalability faults
              based on the synergy between \dcfs and anti-patterns
              (\cref{sec-approach})}.
    \item We evaluate \scalelens on \benchtotal previously
          known real-world
          scalability faults found in \sysca, \syshd, and \sysig, showing the
          effectiveness of \scalelens in detecting and precisely identifying
          root causes for 36 of them. Compared to the baseline, \scalelens improves the \dcf detection by 4.2\texttimes\xspace (\cref{sec-eval-effectiveness}).
    \item We demonstrate the ability of \scalelens to identify \calculate{129+68+137}
          \dcfs with associated anti-patterns in the latest stable
          versions of \sysca, \syshd, and \sysig.
          We discuss their behavior and implications, as well as our ongoing
          process of reporting new faults (\cref{sec-eval-implications}).
\end{itemize}

\section{Empirical Study of Scalability Faults}
\label{sec-empirical-fault-patterns}

This section presents our empirical study on scalability faults, with a
discussion of their root-cause categories and common anti-patterns.
We describe the methodology of our empirical study below.

\label{sec-empirical-methodology}

\subsubsection{System Selection} We \csrewite{considered} \numrepos large-scale
distributed systems: \sysca~\cite{apache_cassandra_web} (CA),
\sysha~\cite{apache_hadoop_web} (HA), \syshb~\cite{apache_hbase_web} (HB),
\syshd~\cite{apache_hdfs_web} (HD), \sysig~\cite{apache_ignite_web} (IG),
\syskf~\cite{apache_kafka_web} (KF), \sysmr~\cite{apache_mapreduce_web} (MR),
\syssp~\cite{apache_spark_web} (SP), \sysst~\cite{apache_storm_web} (ST), and
\sysyr~\cite{apache_yarn_web} (YR).
\csrewite{We chose these systems because 
(1)~they are open-source with public and highly organized JIRA 
issue tracking systems, 
(2)~they are typically deployed at scale and  
(3)~the selection covers the most common
categories of distributed systems, including databases, file systems, 
and parallel computing}.

\subsubsection{Issue Selection} {We collected all (over \totalbugs) issue
reports from the JIRA issue tracking systems of the target systems from 2007 to
2024. Our initial attempt to use keywords to automatically identify reports
describing scalability faults was unfruitful due to the wide variety of
scalability problems and the lack of specific keywords (detailed in
\cref{sec-threats-internal}).  Therefore, we performed a two-pass manual inspection.
In the first pass, we read the title and description of the reports to
identify bug reports that involve the increase in the scale of any aspect of
the system. In the second pass, we performed a more detailed review of the
candidate bug reports by examining the aspects being scaled when symptoms are observed.
We then categorized those into four dimensions: \textit{load}, \textit{data}, \textit{cluster}, and 
\textit{failure}. This process involved 8 individuals and spanned over one year, 
leading to the identification of \numbugs scalability faults, detailed in \cref{tab-dim-systems}.}

\subsubsection{Root-Cause and Anti-Pattern Analysis}
We manually analyzed the \numbugs scalability faults to understand the nature
of each fault and its root cause. This process consisted of examining the
contents of the report, discussions, source code, and available patches/pull requests.
To ensure reliability and minimize bias, we employed a multi-stage adjudication process.
First, each issue was independently reviewed and tagged by two authors, who categorized
it by root-cause category (compute, unbound, bloat, or logic) and its associated anti-pattern.
Second, a third author reviewed all cases where the initial taggers disagreed or
expressed uncertainty. Finally, all authors participated in group discussions to
resolve remaining conflicts and validate the final categorization.
All analyzed issue reports are publicly
available in the respective JIRA repositories, and we provide supplementary material including the complete set of bug reports analyzed in this study,
detailed tagging guidelines, and a full catalog of issue categorizations to support
reproducibility and enable further research.

\begin{figure}[!t]
    \centering
    \includegraphics[width=0.9\columnwidth]{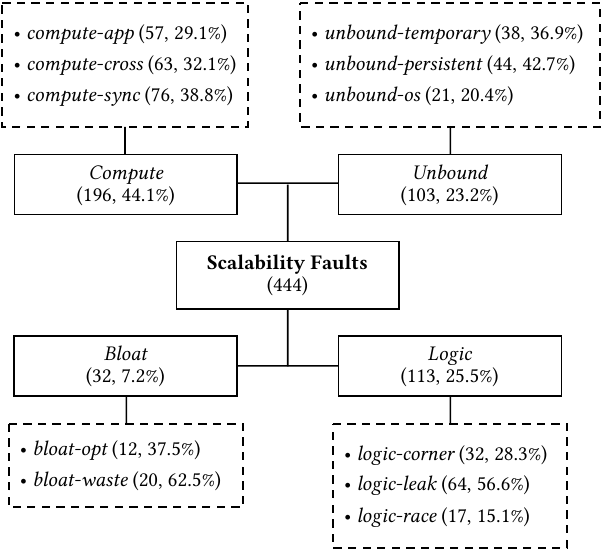}
    \caption{Scalability Fault Anti-Patterns.}
    \label{fig-taxonomy}
\end{figure}

Our analysis has revealed four \textbf{root-cause categories} of scalability
faults, \textit{compute}, \textit{unbound}, \textit{bloat}, and
\textit{logic}, and 11 \textbf{anti-patterns} grouped under them, as shown
in \cref{fig-taxonomy}. Across these anti-patterns, a recurring structural
property stands out: many of them involve code fragments that iterate
(either implicitly or explicitly) a number of times that grows with the
scale of the system. We refer to these as \textbf{dimensional code fragments
(\dcfs)}. \dcfs are not themselves part of the taxonomy;
instead, they are a cross-cutting property shared mainly by the
\textit{compute} and \textit{unbound} anti-patterns. Below, we describe
the root-cause categories and anti-patterns in turn, drawing on \dcfs
where the anti-pattern's behavior is directly tied to scale-dependent
iteration.

\subsection{\textit{Compute} Faults}
\label{sec-rc-compute}

\begin{figure*}
\includegraphics[width=\textwidth]{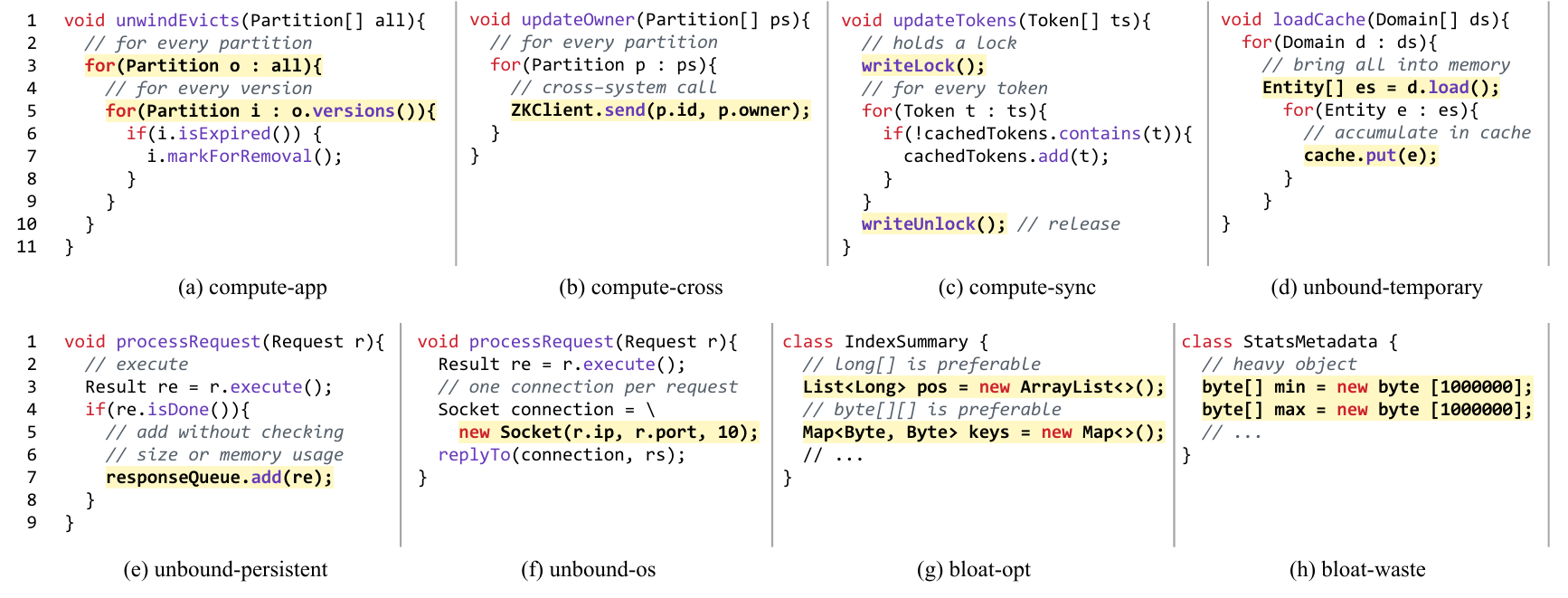}

\caption{Code Samples for the \textit{Compute}, \textit{Unbound}, and \textit{Bloat} Faults. From (a) to (h), they are based on \ignc{8681}, \kfnc{5642}, \canc{5456}, \yrnc{7147}, \canc{15013}, \hanc{15696}, \canc{5506}, and \canc{15400}, respectively.}
\label{fig-rc-samples-all}
\end{figure*}

\textit{Compute} faults are related to bottlenecks caused
by the increase in size of \csrewite{one or more scalable system 
dimensions (\eg \textit{\#partitions} in \cref{fig-rc-samples-all}a}). 
It is the largest category with \textbf{\numcompute} fault reports, 
representing \textbf{\percompute} of the total, and includes $3$ 
anti-patterns:

\csrewite{
$\bullet$ \textbf{compute-app} faults, with \textbf{\numcomputeapp} reports, 
account for \textbf{\percomputeapp} of this category and represent 
cases in which bottlenecks are caused by \dcfs in performance critical paths. 
Even when the computations inside the \dcf 
are not considered costly, the complexity 
(commonly $>=2$) produces notable performance degradation.} 
For example, in \ig{8681} (\cref{fig-rc-samples-all}a), 
\texttt{unwindEvicts} is invoked every time a remote \sysig 
command is processed. This method contains a quadratic \dcf
of $O(P^{2}$, $P = \text{\textit{\# partitions}})$ 
(highlighted at lines $3$ and $5$).
Since remote command execution is part of the critical path, 
invoking this method is reportedly problematic.

\csrewite{$\bullet$ \textbf{compute-cross} faults, 
with \textbf{\numcomputecross} 
reports, account for \textbf{\percomputecross} of this category and 
represent cases in which bottlenecks are caused by \dcfs containing 
cross-system (\eg methods from pluggable components, external clients, 
or \io operations) method calls. 
When cross-system method calls are performed inside \dcfs, 
their computational cost gets amplified 
\citep{chatyiomicrosoft.web, DBLP:conf/icse/YangSLYC18}
by the size of the related dimension, reportedly leading to 
severe performance degradation.}
For example, in \kf{5642} (\cref{fig-rc-samples-all}b), 
\texttt{updateOwner} contains a linear \dcf that 
iterates over each partition to update the ownership for a \syskf topic. 
Each iteration triggers synchronous communication 
with a cross-layer component (\syszk, highlighted at line $5$). 
As the number of partitions grows, the number of synchronous messages 
increases and the execution time of the whole operation 
becomes dependent on the network speed and the load on the external component.

\csrewite{$\bullet$ \textbf{compute-sync} faults, 
with \textbf{\numcomputesync} reports, account for 
\textbf{\percomputesync} of this category and represent cases 
in which synchronization bottlenecks are caused by \dcfs protected 
by global locks. 
Such fault anti-patterns have been observed
previously \citep{DBLP:conf/icmla/LiscanoARASC23, DBLP:conf/fast/StuardoLSKLCLG19} 
in centralized architectures such as \syshd and include both cases 
in which the locks wrap \dcfs and cases 
in which the locks are held inside the \dcf, 
the former being the most frequent.}
For example, in \ca{5456} (\cref{fig-rc-samples-all}c), 
the method \texttt{updateTokens} is invoked every 
time \textit{range movements} are performed to maintain 
consistency when a \sysca cluster's topology is changing.
The operation contains a linear \dcf (line $4$) 
and is performed while holding a global lock (highlighted at line $2$). 
Other threads, in particular the ones that handle membership changes, 
have to wait until the whole operation finishes. As the number of 
tokens grows, the execution time of the code block between 
lines $4$ and $8$ grows too, reportedly causing a 
negative impact on elasticity.

\subsection{\textit{Unbound} Faults}
\label{sec-rc-unb}

\textit{Unbound} faults are related to unconstrained resource consumption 
caused by the increase in size of \csrewite{one or more scalable system 
dimensions (\eg \textit{\#entities/domains} in \cref{fig-rc-samples-all}d}).
This category, with \textbf{\numunb} fault reports, 
represents \textbf{\perunb} of the total and includes $3$ anti-patterns:

\csrewite{$\bullet$ \textbf{unbound-temporary} faults, with
\textbf{\numunballoc} reports, account for \textbf{\perunballoc} of
this category and represent cases in which 
temporary memory allocations, 
such as stack-level data structures, grow without bounds in 
response to the growth of one or more dimensions, 
reportedly causing out-of-memory errors and performance 
degradation due to frequent garbage collection.
This anti-pattern can be intuitively related to the
lack of buffering or paging when loading
data from external sources (\eg files or databases) 
\citep{exfetchingmicrosoft.web, DBLP:conf/icse/ChenSJHNF14}.}
For example, in \yr{7147} (\cref{fig-rc-samples-all}d), the method
\texttt{loadCache} is invoked every time the component
\textit{Timeline Server} \citep{yarnats.web} of 
\sysyr is started and contains a quadratic \dcf
of $O(D*E$, $D = \text{\textit{\# domains}}, E = \text{\textit{\# entities}})$.
The cache is loaded from disk (highlighted at line $7$), 
but the whole file is materialized in memory in one method call 
(highlighted at line $4$), causing the component to run out 
of memory when the related files get too big. 
As reported, the component is killed and it can be restarted 
only if there is more memory available or the caching feature is disabled.

\csrewite{
$\bullet$ \textbf{unbound-persistent} faults, with
\textbf{\numunbcollection} reports, account for
\textbf{\perunbcollection} of this category and represent cases 
in which long-lived data structures, such as 
inbound/outbound message processing queues or multi-purpose caches, 
grow without bounds in response to the growth of one or more dimensions, 
reportedly causing out-of-memory errors. 
Note that the difference between this and the previous category 
is related to object lifetime, purpose and expected functionality: 
While the previous refers to temporary allocations typically used 
as containers/helpers, this category refers to 
persistent allocations that, accompanied by the necessary logic, 
are supposed to accommodate the growth of one or more dimensions 
but fail to do so due to issues in said logic.}
For example, in \ca{15013}
(\cref{fig-rc-samples-all}e), the \sysca request processor contains 
a \dcf of $O(R$, $R = \text{\textit{\# requests}})$ that accumulates reportedly 
large responses in a persistent queue (highlighted at line $7$). 
If the responding threads (which consume those objects) cannot
keep up, the queue grows without bounds and the node runs out
of memory.

\csrewite{
$\bullet$ \textbf{unbound-os} faults, with \textbf{\numunbos} issues,
account for \textbf{\perunbos} of this category and represent cases 
in which \os resource allocations, such as threads and file descriptors, 
grow without bounds in response to the growth of one or more dimensions 
potentially exhausting those resources.
These faults are uncommon in newer versions since the unbounded
increase of threads and/or sockets is a known issue and is typically
targeted in early fixes.}
For example, in \ha{15696} (\cref{fig-rc-samples-all}f), an
encryption-related \syshd component contains 
a \dcf of $O(R$, $R = \text{\textit{\# requests}})$
that creates a single \texttt{Socket}
(with a corresponding file descriptor) every time a request is
processed. Even if these connections are short-lived, the default
idle-timeout is set to $10$ seconds (highlighted at line $5$). 
As reported, an explosive increase in the number of requests ends 
up in failures (the system runs out of file descriptors).

\subsection{\textit{Bloat} Faults}
\label{sec-rc-bloat}

\csrewite{\textit{Bloat} faults are related to data
structures with scalability-limiting
design \cite{DBLP:conf/sigsoft/Soto-ValeroDB21,
DBLP:journals/software/MitchellSS10,
DBLP:journals/tosem/NguyenWBFX18}, 
causing unexpected increases in memory consumption at larger scales. 
This category, with \textbf{\numbloat} fault reports, represents
\textbf{\perbloat} of the total and includes $2$ anti-patterns:}

\csrewite{
$\bullet$ \textbf{bloat-opt} faults, with \textbf{\numbloatopt}
issues, account for \textbf{\perbloatopt} of this category 
and represent cases in which data structure design includes 
``space-time trade-offs'' \cite{DBLP:conf/sigmod/DayanI18, 
DBLP:conf/icde/ZhangWQWOHLF024} that, albeit intended to 
improve performance, obviate memory constraints at larger scales.
For example, in \ca{5506} (\cref{fig-rc-samples-all}g), \sysca nodes
utilize a data structure called \texttt{IndexSummary} that
uses the collections framework \citep{javacollections.web} to store 
internal indexes. As the number of live \texttt{IndexSummary} instances grows, a $2 \times$ per-element overhead in positions
(highlighted at line $3$) and a $10 \times$ per-element overhead 
in keys causes $70$\% memory bloat in a cluster with billions of rows.}

\csrewite{$\bullet$ \textbf{bloat-waste} faults, 
with \textbf{\numbloatwaste} reports, 
account for \textbf{\perbloatwaste} of this category and 
represent cases in which data structure design includes 
unnecessarily large preallocated fields 
(\eg fixed arrays or buffers), that obviate memory 
constraints at larger scales.
Note that the difference between this and the previous category 
is related to functionality: The previous focuses on allocations 
related to more complex techniques (such as indexing) which require a much 
more fine-grained design, while this category focuses 
on unnecessarily large (and often unused) space being preallocated 
(\ie \textit{wasted}). 
For example, in \ca{15400} (\cref{fig-rc-samples-all}h), 
\sysca creates one \texttt{StatsMetadata} instance 
for each \textit{BigTableReader}, an internal data structure 
whose size is correlated to the load dimension. The former declares 
two fixed-size $1$MB buffers (highlighted at lines $3$ and $4$) 
that, according to developers, are frequently not used in their entirety. 
When under high
load, many live \textit{BigTableReader} instances lead to high memory
usage and eventually end in out-of-memory errors, affecting not one
but several nodes in the cluster.}
\subsection{\textit{Logic} Faults}
\label{sec-rc-logic}

\begin{figure*}[ht]
    \centering
    \includegraphics[width=\textwidth]{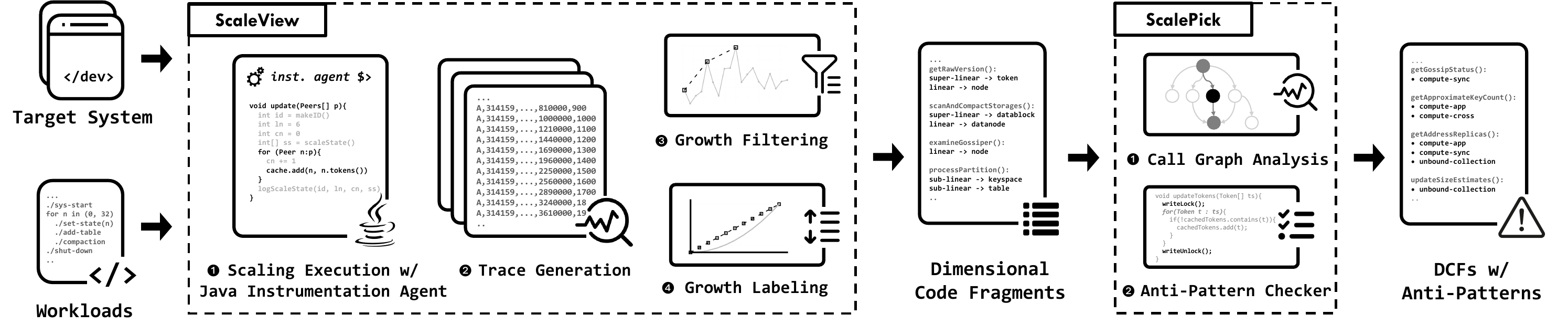}
    \caption{\scalelens Workflow.}
    \label{fig-scaleview}
\end{figure*}

\textit{Logic} faults are related to logic issues that become 
visible at large scales, such as corner cases, memory or resource leaks, 
and data races. This category, with \textbf{\numlogic} fault reports, 
represents a surprising \textbf{\perlogic} of the total and 
includes $3$ anti-patterns:

\csrewite{
$\bullet$ \textbf{logic-corner} faults, with \textbf{\numlogiccorner} 
issues, account for \textbf{\perlogiccorner} of this category 
and represent cases in which faulty logic (\eg 
wrong choice of datatypes) is problematic at larger scales but 
harmless at smaller scales.
For example, in \st{3256}, a large \sysst topology is 
used to communicate with custom terminals in a 
client-server fashion. The communication is handled 
by \textit{worker} threads, which are uniquely identified 
across the whole deployment using a numeric 16-bit unsigned integer. 
Due to this, no more than $32$K workers can be instantiated concurrently, 
which imposes a severe limit to parallelism 
(in the reporter's words, 
``\textit{... is a disaster for large scale computing}'').}

\csrewite{
$\bullet$ \textbf{logic-leak} faults, with \textbf{\numlogicleak} issues, 
account for \textbf{\perlogicleak} of this category and represent cases 
in which memory, \os resources, or storage leaks at larger scales.} 
For example, in \hd{13039}, \syshd \textit{erasure coding} 
component exhibits a classic \os (file descriptor) 
leak: \texttt{createReader} is invoked when 
creating a \texttt{Reader} object on a per-request basis. 
However, a connection is created but never closed in the case of 
file creation failures. \csrewite{This leads to extended 
downtimes (no more file descriptors available) when many of 
these failures happen.}

\csrewite{
$\bullet$ \textbf{logic-race} faults, with \textbf{\numlogicrace} 
issues, account for \textbf{\perlogicrace} of this category and 
represent cases in which race conditions are observed at larger scales 
(\eg when a large number of peers join the cluster 
at the same time), causing effects that range 
from data inconsistencies to system failures.}
For example, in \ha{16385}, a $53$K-container 
heavily loaded \sysha \syshd cluster is dealing with multiple 
node failures at the same time. In the middle of this 
failure storm, a race condition related to the calculation 
of the number of live peers triggers an assertion failure 
that ends up killing the \syshd \textit{namenode}.

\subsection{Key Insights}
\label{sec-empirical-final-remarks}

Our study has confirmed that scalability faults are challenging to
detect. We identified four main categories of scalability faults, from
which \textit{compute} and \textit{unbound} stand out; more than half
of all faults (\ratio{299}{444}) fall into these categories, which are
caused by the combination of \dcfs and anti-patterns. From them, we
observe the following scalability anti-patterns:
\begin{itemize}
    \item \textit{compute-app}: \dcfs placed in performance-critical paths.
    \item \textit{compute-cross}: \dcfs with API or IO operations.
    \item \textit{compute-sync}: \dcfs protected by global locks.
    \item \textit{unbound-temporary}: \dcfs with memory allocations.
    \item \textit{unbound-persistent}: \dcfs that grow the size of objects.
    \item \textit{unbound-os}: \dcfs that consume OS resources.
\end{itemize}

Based on these observations, identifying \dcfs and associated
anti-patterns is essential to detect scalability faults. We also
observe that:

$\bullet$ \textbf{The nature of \dcfs} can be \textbf{explicit} or
\textbf{implicit}. Explicit \dcfs are \texttt{for} or \texttt{while} loops
directly visible in the source code of the system or its related libraries,
as shown in Figures \ref{fig-rc-samples-all}a-d, and implicit \dcfs refer to
cases where no loops are directly present, but the number of method
invocations increases with one or more system dimensions as shown in Figures
\ref{fig-rc-samples-all}e-f. Due to the nature of \dcfs, especially implicit
ones, they are difficult to detect without runtime information.

$\bullet$ \textbf{\dcfs are not necessarily problematic by themselves},
as they represent basic building blocks of distributed systems.
For example, for \textit{compute}-related \dcfs, we observe that
they become problematic only when present in performance
sensitive paths, make external API or \io calls, or are protected by global locks. Therefore, detecting \dcfs alone is not sufficient and identifying associated anti-patterns is essential to accurately detect scalability faults.

$\bullet$ \textbf{\dcf complexity matters}, but it is not a strong
indicator of faults by itself. For example, high-complexity functions
could be considered acceptable in background paths but problematic in
foreground or user-facing paths.

In addition to \textit{compute} and \textit{unbound} faults, we also
observe \textit{bloat} and \textit{logic} faults. \textit{Bloat}
faults are related to inefficient data structure designs that lead to
unnecessary memory overhead as the system scales, while \textit{logic}
faults are logic issues that become visible when the system is
deployed at larger scales. Unlike \textit{compute} and
\textit{unbound} faults, these two categories do not involve \dcfs,
and may require memory footprint and concurrency analyses
for their detection. Together, they account for a smaller portion of
the scalability faults (\ratio{145}{444}).

In the next section, we describe our approach, \scalelens, to detect
\textit{compute} and \textit{unbound} scalability faults. We leave
further analysis of \textit{bloat} and \textit{logic} faults as future
work.

\section{Our Approach: \scalelens}
\label{sec-approach}

We present \scalelens, an approach for detecting scalability faults in
large-scale distributed systems. \scalelens has two major components, illustrated in \cref{fig-scaleview}:
(1) \scaleview that performs \textit{dynamic} analysis to identify
\dcfs, and (2) \scalepick that performs \textit{static} analysis
to detect anti-patterns associated with \dcfs.

\subsection{\scaleview}

The goal of \scaleview is to list all \dcfs for a given system and
their relationships with the system dimensions. Such information can
help developers to \textit{view} all potential scalability bottlenecks
from a scaling perspective prior to actual production deployment.

Given the source code of a target system and a set of scaling workloads, 
\scaleview \textbf{(1)} instruments the code to obtain
execution traces, \textbf{(2)} parses and analyzes the execution
traces to discover execution trends, \textbf{(3)} filters out noise to
identify real growth, and \textbf{(4)} labels the \dcfs according to
their relationships to the system dimensions.

\subsubsection{Scaling Workloads and System Dimensions}
\label{sec-approach-workload}

The first step is to write a \textbf{scaling workload}: a script that
systematically increases one or more scalable system dimensions (the
\textit{scalable} aspects of interest) while invoking the target system's
APIs. This process is manual and requires familiarity with deployment
and operational practices of the target system. It is a one-time effort
and can be reused across different versions of the system (implications
are discussed in \cref{sec-threats-external}).

\begin{figure}
        \centering
        \includegraphics[width=0.95\columnwidth]{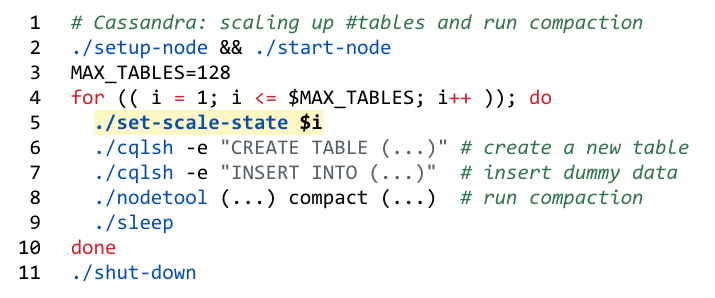}
    \caption{Scaling Workload Example.}
    \label{fig-workload}
\end{figure}

\cref{fig-workload} shows an example workload scaling the
number of tables in a \sysca cluster. The maximum number of tables is
set on line 3, and the \texttt{for} loop on line 4 adds one table at a
time. On lines 6 and 7, a table is created and data is
inserted via \texttt{cqlsh}, the database operation interface provided
by \sysca. On line 8, compaction is run via \texttt{nodetool}, the
system management utility provided by \sysca. Compaction is optional,
and can be replaced with other system operations (\eg snapshot, scrub,
repair, \etc) based on the interest of the user of \scaleview.

The function \texttt{set-scale-state} on line $5$ of \cref{fig-workload} is
an API provided by \scaleview for reporting the current \textbf{scale
state} of the workload: the numeric value(s) of the dimension(s) being
scaled at that moment of execution (here, the current number of tables,
\eg $17$ after the 17th loop iteration). \scaleview uses this reported
value to correlate each observed iteration count of a code fragment with
the scale of the target dimensions. A multi-dimensional scale state is
also supported: for example, a workload that scales tables and rows
simultaneously would call \texttt{set-scale-state} with a pair of values,
such as \texttt{(17, 10000)} once the 17th table contains 10000 rows.

\subsubsection{Runtime Instrumentation}
\label{sec-approach-instrumentation}

After scaling workloads are triggered, 
\scaleview monitors the relationship between the 
current size of the dimensions and the number of iterations performed by a code fragment.
Such monitoring is done by using bytecode-level runtime instrumentation 
\cite{javaassist-github.web, javainstrumentation.web}.
Based on the insights from \cref{sec-empirical-fault-patterns}, 
\scaleview considers both explicit and implicit iterations.

\begin{figure}
  \includegraphics[width=\columnwidth]{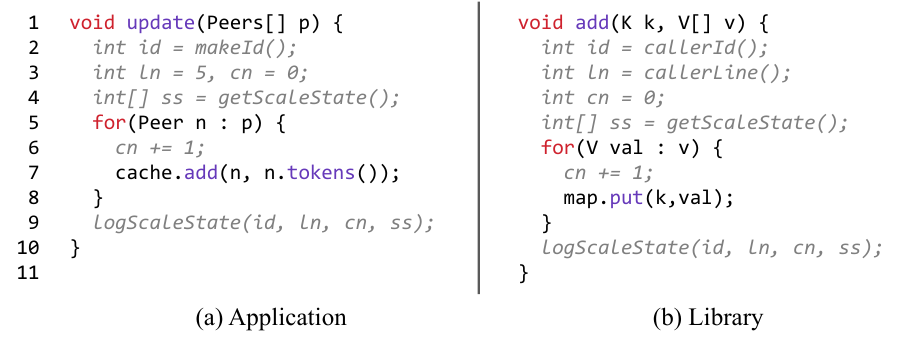}
  \caption{Runtime Instrumentation. Code in grey is instrumented.}
  \label{fig-instrumentation}
\end{figure}

$\bullet$ \textbf{Explicit.} As shown in \cref{fig-instrumentation}a, once a 
loop is detected by \scaleview in the source code of the target system, 
\scaleview adds a unique ID, line number, and a counter variable to the loop.  
Library loops are also instrumented but in this case the unique ID corresponds 
to the ID of the caller, as shown in \cref{fig-instrumentation}b. This is done to maintain the scope of 
reporting within the target system. Finally, for both types of loops, \scaleview 
retrieves the scale state information with \texttt{getScaleState()} and logs the scale state along 
with the counter variables with the \texttt{logScaleState()}.

$\bullet$ \textbf{Implicit.} \scaleview captures
implicit iterations by capturing the relationship between 
the number of invocations of each method and the current state of the scaling dimensions.

\subsubsection{Trace Generation}
\label{sec-approach-analyzing}

The combination of scaling workloads and instrumentation produces 
traces that capture runtime information about loop executions. 
Specifically, a tuple $\langle
type, methodID, line\#, execID, \#iterations, \mathcal{SS} \rangle$ is
recorded for each loop execution. In the tuple, $type \in \{A, L, I\}$
represents the type of the loop: application, library, or implicit. $methodID$ and $line\#$ 
record the method and location of the loop, respectively. $execID$ is 
a unique identifier for each execution of the loop, and $\#iterations$ is the final 
iteration count of a given loop execution. $\mathcal{SS}$ represents the 
scale state in an array with one or more numbers for one- or multi-dimensional 
executions, respectively. For example, tuple $\langle
A,314159,64,265358,810000,900 \rangle$ describes the loop execution with ID $265358$: 
an execution of the application loop located in the method with ID $314159$ on line $64$ iterated a total of
$810000$ times when the number of nodes (\ie the scale state)
reached $900$, respectively.

\subsubsection{Growth Filtering}
\label{sec-approach-filtering}

\begin{figure}
    \begin{subfigure}[b]{0.24\columnwidth}
        \centering
        \includegraphics[width=\textwidth]{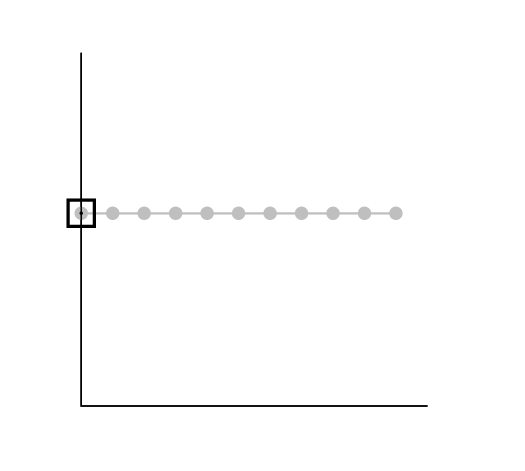}
        \caption{clear flat}
        \label{fig-growth-flat}
    \end{subfigure}
    \begin{subfigure}[b]{0.24\columnwidth}
        \centering
        \includegraphics[width=\textwidth]{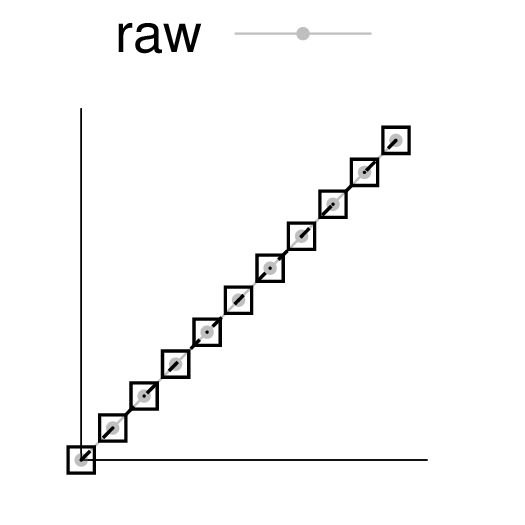}
        \caption{clear growth}
        \label{fig-growth-growth}
    \end{subfigure}
    \begin{subfigure}[b]{0.24\columnwidth}
        \centering
        \includegraphics[width=\textwidth]{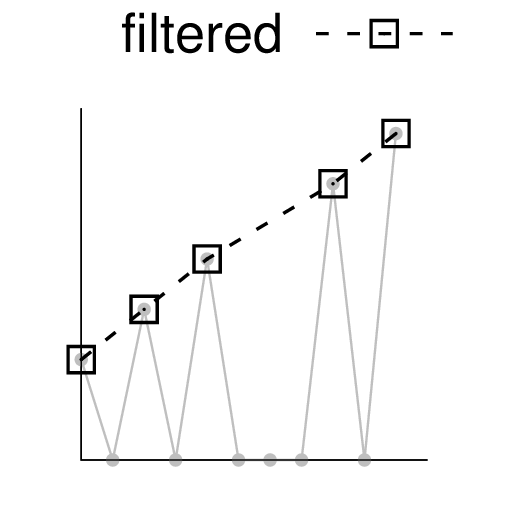}
        \caption{noisy growth}
        \label{fig-growth-noise-growth}
    \end{subfigure}
    \begin{subfigure}[b]{0.24\columnwidth}
        \centering
        \includegraphics[width=\textwidth]{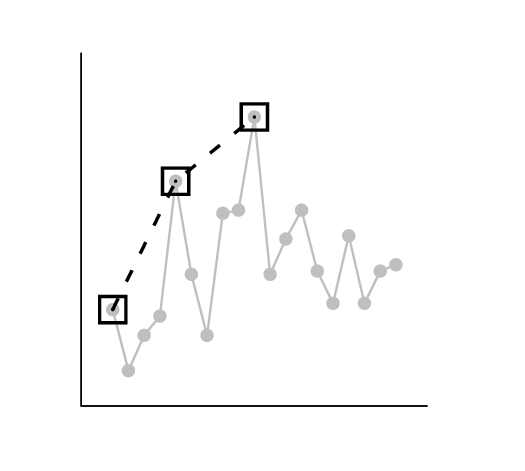}
        \caption{noisy flat}
        \label{fig-growth-noise-flat}
    \end{subfigure}
    \caption{Growth Patterns.}
\label{fig-growth-noise}
\end{figure}

The goal of growth filtering is to detect \dcfs by identifying iterative code 
fragments with a positive correlation between the number of iterations and the target scaling dimensions, 
and discarding the rest. This is done with heuristics based on the growth 
patterns shown in \cref{fig-growth-noise}. Besides the trivial clear 
flat (\cref{fig-growth-flat}) and clear growth (\cref{fig-growth-growth}) 
patterns, we also observed noisy growth (\cref{fig-growth-noise-growth}) 
and noisy flat (\cref{fig-growth-noise-flat}) patterns. \scaleview 
performs a 3-step empirical-driven filtering process: \textbf{(1)} only keep the 
data points that are larger than the maximum of all the previous data points (\ie the dashed boxes in \cref{fig-growth-noise}), 
\textbf{(2)} remove the code fragments with less than $10\%$ of data points retained after the first step 
(\eg \cref{fig-growth-flat} and \cref{fig-growth-noise-flat}), and \textbf{(3)} keep the code fragments whose correlation 
to the scaling dimension is high (\ie $\geq 0.9$). Such a filtering process excludes both clear and noisy flat patterns.

\subsubsection{Growth Labeling}
\label{sec-approach-ranking}

Finally, \scaleview categorizes \dcfs based on their previously 
filtered growth trends as (1) super-linear, (2) linear, or (3) sub-linear. 
To do so, we select a set of theoretical complexity models 
(\eg $I = O(D^2)$ is super-linear, $I = O(D)$ for linear, and $I = O(\log{D})$ 
for sub-linear, where $I$ is the number of iterations and $D$ is 
the scale of dimension) and use similarity measures 
\cite{DBLP:conf/esa/AronovHKWW06} to detect the closest theoretical 
model and assign the corresponding label to the fragment.
We consider that this simple labeling can be useful 
for developers to estimate the urgency or priority of the issue.
For example, a super-linear (\eg $O(N^2)$ or above) code 
fragment intuitively brings more negative impacts 
than a linear (\eg$O(N)$) fragment under the same condition.

\subsection{\scalepick}
\label{section-static-checker}

\begin{table*}[]
    \centering
    \caption{Evaluation Results. $T_{DCF}$ denotes the type of reported \dcf, where \textit{E} stands for explicit 
    and \textit{I} for implicit. In \textit{DCF} columns for both \scalelens and \sfind, \gxmark\xspace denotes that the related \dcf was not detected and \cmark\xspace denotes positive results. In \textit{Anti-Pattern} column under \scalelens, \xmark\xspace denotes the anti-pattern is not detected given the detection of \dcf, and empty cell means not evaluated because the \dcf is not detected.}
    \scriptsize
    \setlength{\tabcolsep}{.2pt}
    \begin{adjustbox}{width=\textwidth}
    \begin{tabular}{l@{\hskip 4pt}c@{\hskip 4pt}r@{\hskip 7pt}c@{\hskip 4pt}c@{\hskip 4pt}c@{\hskip 5pt}V{2}@{\hskip 5pt}l@{\hskip 4pt}c@{\hskip 4pt}r@{\hskip 7pt}c@{\hskip 4pt}c@{\hskip 4pt}c@{\hskip 4pt}V{2}@{\hskip 5pt}l@{\hskip 4pt}c@{\hskip 7pt}r@{\hskip 4pt}c@{\hskip 4pt}c@{\hskip 4pt}c}
        \toprule
        \multicolumn{3}{c}{Issue Information}                        & \multicolumn{2}{c}{\scalelens\phantom{S}}  & \sfind & \multicolumn{3}{c}{Issue Information}                             & \multicolumn{2}{c}{\scalelens\phantom{S}}               & \sfind                         & \multicolumn{3}{c}{Issue Information}                                  & \multicolumn{2}{c}{\scalelens\phantom{S}}  & \sfind \\
        \cmidrule(lr{10pt}){1-3} \cmidrule(lr{10pt}){4-5} \cmidrule(l{5pt}r{15pt}){6-6} \cmidrule(lr{10pt}){7-9} \cmidrule(lr{10pt}){10-11} \cmidrule(l{5pt}r{15pt}){12-12} \cmidrule(lr{10pt}){13-15} \cmidrule(lr{10pt}){16-17} \cmidrule(l{5pt}r{5pt}){18-18}
        ID           & $T_{DCF}$            & Anti-Pattern                & DCF     & Anti-Pattern              & DCF        & ID           & $T_{DCF}$                 & Anti-Pattern                & DCF                  & Anti-Pattern              & DCF                                & ID                     & $T_{DCF}$            & Anti-Pattern                & DCF     & Anti-Pattern              & DCF        \\
        \midrule
        \canc{19534} & \textit{I}      & \textit{u-persistent} & \gxmark & \phantom{\cmark} & \gxmark    & \canc{14660} & \textit{E \& I}      & \textit{c-sync}       & \cmark               & \cmark               & \gxmark                            & \hdnc{16100}           & \textit{E}      & \textit{c-sync}       & \cmark  & \cmark               & \cmark     \\
        \canc{19477} & \textit{E \& I} & \textit{c-cross}      & \cmark  & \cmark               & \gxmark    & \canc{14239} & \textit{E}           & \textit{u-persistent} & \cmark               & \cmark               & \gxmark                            & \hdnc{15621}           & \textit{I}      & \textit{u-persistent} & \gxmark & \phantom{\cmark} & \gxmark    \\
        \canc{19412} & \textit{E \& I} & \textit{u-persistent} & \cmark  & \cmark               & \gxmark    & \canc{14096} & \textit{E}           & \textit{u-persistent} & \cmark               & \cmark               & \gxmark                            & \hdnc{15406}           & \textit{E}      & \textit{c-sync}       & \cmark  & \cmark               & \gxmark    \\
        \canc{19336} & \textit{I}      & \textit{u-temporary} & \cmark  & \cmark               & \gxmark    & \canc{13993} & \textit{I}           & \textit{c-cross}      & \cmark               & \cmark               & \gxmark                            & \hdnc{14859}           & \textit{E}      & \textit{c-app}        & \cmark  & \cmark               & \cmark     \\
        \canc{19107} & \textit{E \& I} & \textit{c-app}        & \cmark  & \cmark               & \gxmark    & \canc{13948} & \textit{E}           & \textit{c-sync}       & \cmark               & \cmark               & \cmark                             & \hdnc{14854}           & \textit{E}      & \textit{c-sync}       & \gxmark & \phantom{\cmark} & \gxmark    \\
        \canc{18773} & \textit{E \& I} & \textit{c-app}        & \cmark  & \cmark               & \gxmark    & \canc{13923} & \textit{E \& I}      & \textit{c-sync}       & \cmark               & \cmark               & \gxmark                            & \hdnc{14771}           & \textit{E}      & \textit{c-cross}      & \gxmark & \phantom{\cmark} & \gxmark    \\
        \canc{18546} & \textit{E \& I} & \textit{u-persistent} & \cmark  & \cmark               & \gxmark    & \canc{13569} & \textit{I}           & \textit{u-persistent} & \gxmark              & \phantom{\cmark} & \gxmark                            & \hdnc{14657}           & \textit{I}      & \textit{c-sync}       & \cmark  & \cmark               & \gxmark    \\
        \canc{17787} & \textit{I}      & \textit{u-temporary} & \cmark  & \cmark               & \gxmark    & \canc{13299} & \textit{I}           & \textit{u-persistent} & \gxmark              & \phantom{\cmark} & \gxmark                            & \hdnc{14613}           & \textit{E}      & \textit{c-app}        & \cmark  & \cmark               & \cmark     \\
        \canc{17691} & \textit{E \& I} & \textit{c-sync}       & \cmark  & \cmark               & \gxmark    & \canc{13215} & \textit{E \& I}      & \textit{c-app}        & \cmark               & \cmark               & \gxmark                            & \hdnc{14370}           & \textit{E}      & \textit{c-cross}      & \gxmark & \phantom{\cmark} & \gxmark    \\
        \canc{17342} & \textit{E \& I} & \textit{c-app}        & \cmark  & \cmark               & \gxmark    & \canc{13065} & \textit{E}           & \textit{c-sync}       & \gxmark              & \phantom{\cmark} & \gxmark                            & \hdnc{14366}           & \textit{E}      & \textit{c-sync}       & \cmark  & \cmark               & \cmark     \\
        \canc{16380} & \textit{E \& I} & \textit{c-sync}       & \cmark  & \cmark               & \gxmark    & \canc{12281} & \textit{I}           & \textit{c-sync}       & \cmark               & \cmark               & \gxmark                            & \hdnc{14201}           & \textit{I}      & \textit{c-cross}      & \gxmark & \phantom{\cmark} & \gxmark    \\
        \canc{16261} & \textit{I}      & \textit{u-persistent} & \gxmark & \phantom{\cmark} & \gxmark    & \canc{12245} & \textit{E}           & \textit{c-sync}       & \gxmark              & \phantom{\cmark} & \gxmark                            & \hdnc{14171}           & \textit{E}      & \textit{c-app}        & \cmark  & \cmark               & \cmark     \\
        \canc{16201} & \textit{E}      & \textit{u-temporary} & \cmark  & \cmark               & \gxmark    & \canc{11748} & \textit{I}           & \textit{u-persistent} & \gxmark              & \phantom{\cmark} & \gxmark                            & \hdnc{13821}           & \textit{I}      & \textit{c-sync}       & \gxmark & \phantom{\cmark} & \gxmark    \\
        \canc{15364} & \textit{I}      & \textit{c-app}        & \cmark  & \cmark               & \gxmark    & \canc{10654} & \textit{E}           & \textit{c-sync}       & \gxmark              & \phantom{\cmark} & \gxmark                            & \hdnc{13768}           & \textit{E}      & \textit{c-sync}       & \gxmark & \phantom{\cmark} & \gxmark    \\
        \canc{15141} & \textit{E \& I} & \textit{c-sync}       & \cmark  & \cmark               & \gxmark    & \canc{9882}  & \textit{E \& I}      & \textit{c-sync}       & \cmark               & \cmark               & \cmark                             & \hdnc{13692}           & \textit{E}      & \textit{u-persistent} & \cmark  & \cmark               & \cmark     \\
        \canc{15013} & \textit{I}      & \textit{u-persistent} & \gxmark & \phantom{\cmark} & \gxmark    & \hdnc{17097} & \textit{I}           & \textit{c-sync}       & \cmark               & \cmark               & \gxmark                            & \ignc{14076}           & \textit{E \& I} & \textit{u-persistent} & \cmark  & \cmark               & \gxmark    \\
        \canc{14855} & \textit{E \& I} & \textit{u-persistent} & \cmark  & \cmark               & \gxmark    & \hdnc{17096} & \textit{E}           & \textit{u-temporary} & \cmark               & \xmark               & \cmark                             & \ignc{12189}           & \textit{I}      & \textit{c-cross}      & \cmark  & \cmark               & \gxmark    \\
        \canc{14840} & \textit{I}      & \textit{u-persistent} & \cmark  & \xmark               & \gxmark    & \hdnc{16989} & \textit{I}           & \textit{c-sync}       & \cmark               & \cmark               & \gxmark                            & \ignc{12087}           & \textit{I}      & \textit{c-sync}       & \cmark  & \cmark               & \gxmark    \\
        \cline{7-18}
        \canc{14747} & \textit{I}      & \textit{u-persistent} & \gxmark & \phantom{\cmark} & \gxmark    &              & \phantom{\cmark} & \phantom{\cmark}        & \phantom{\cmark} & \phantom{\cmark} & \raisebox{-0.5ex}{\textit{Total}} & \raisebox{-0.5ex}{55} &                 &     \phantom{\cmark}    & \raisebox{-0.5ex}{38}      & \raisebox{-0.5ex}{36}                   & \raisebox{-0.5ex}{9}          \\
        \bottomrule
\end{tabular}
\end{adjustbox}
\label{tab-eval-full}
\end{table*}

\dcfs reported by \scaleview represent potential scalability
bottlenecks in the system, but 
as discussed previously, not all 
\dcfs are necessarily problematic. 
Thus, given the source code of a target system, as 
well as the \dcfs found by \scaleview, 
\scalepick \textbf{(1)} performs call graph
analysis from \dcfs, and \textbf{(2)} 
uses static checkers to identify 
associated anti-patterns, if any.

\scalepick incorporates a set of simple static
checkers that identify the first $6$ anti-patterns from our
empirical study using \codeql~\citep{codeql.web}, and
all checkers are applied to each \dcf to flag
associated anti-patterns. \scalepick's checkers use \textbf{program slicing}
\citep{DBLP:journals/ac/BinkleyG96}, a technique that computes the set of
statements that may influence (backward slice) or be influenced by
(forward slice) a given statement. We apply slicing over the call graph
of the target system: a backward slice from a \dcf gathers the methods
that can reach it, while a forward slice from a \dcf gathers the methods
and the allocation, I/O, or synchronization sites it may reach. Each
checker below inspects a specific slice and flags the \dcf if the slice
contains a checker-specific sink. Details
of the checkers are discussed below.

\boldbullet{compute-app} checker detects \dcfs in performance or
operational critical paths. Those code paths could be
user-facing or foreground depending on the testing requirements. 
To detect this anti-pattern, we perform backward 
slicing~\citep{DBLP:journals/ac/BinkleyG96} from the \dcfs 
over the call graph to check reachability from any
specified user-facing or foreground API calls. 
Reachable \dcfs are flagged as \textit{compute-app}.

\boldbullet{compute-cross} checker finds external API or \io
operations within \dcfs. Such operations 
could be read/write interactions with disk, 
network, or database, which are often performed in a
synchronous fashion. To detect this anti-pattern, we perform forward
slicing~\citep{DBLP:journals/ac/BinkleyG96} from the \dcfs over the
call graph to find \dcfs associated with external API or \io
invocations. Such \dcfs are then flagged as \textit{compute-cross}.

\boldbullet{compute-sync} checker finds instances where lock
contentions may be caused by \dcfs holding a global lock. Such lock
contentions typically happen in foreground/background interactions.
To detect this anti-pattern, we perform 
both backward and forward slicing from the
\dcfs over the call graph. If a \dcf holds a global lock that is also
used by other components, then it is flagged as \textit{compute-sync}.

\boldbullet{unbound-temporary} checker finds unbounded memory allocations
whose lifetime is local to the method that contains the \dcf. We
compute the forward slice from the \dcf and check whether the slice
contains a memory-growth sink on a \emph{method-local} target, \ie an
object scoped to the method: (i) a
fresh allocation (\eg a constructor call for a Java
collection class such as \texttt{ArrayList}, \texttt{HashMap}, or
\texttt{StringBuilder}; a Java array creation \texttt{new T[n]}; or an
external buffer-returning API whose return value is not released in the
slice), or (ii) a size-growing call (\eg \texttt{append}, \texttt{add},
\texttt{put}) on a mutable container declared in the containing method
(and hence unreachable after the method returns). If either holds, the
\dcf is flagged as \textit{unbound-temporary}.

\boldbullet{unbound-persistent} checker detects unbounded growth of
long-lived objects reachable from a \dcf, \ie objects whose lifetime
outlives the enclosing method. We compute the forward slice from the
\dcf and check whether it contains a size-growing call on a
\emph{non-local} target: an instance field, a static field, or a
mutable object passed by reference (\eg a shared queue, map, or
cache). If such a call is in the slice, the \dcf is flagged as
\textit{unbound-persistent}.

\boldbullet{unbound-os} checker identifies \dcfs that consume system
resources, such as file descriptors, sockets, or threads, without
proper management. Such resources are often limited and could be
exhausted when the system scales. To detect this anti-pattern, we
perform forward slicing from the \dcfs over the call graph to see if
the \dcf consumes system resources. If so, the \dcf is flagged as
\textit{unbound-os}.

To demonstrate these checkers in practice, consider the for-loop \dcf
from \yr{6188} (the motivating example from~\cref{sec:intro}). In this
example, a \texttt{StringBuilder} declared within the method
containing the \dcf is updated via repeated calls to \texttt{append},
which is reachable from the \dcf. Because the buffer's size grows with
every iteration, the unbound-temporary checker identifies this as a
``size-growth sink'' under case (ii). This detection aligns with both
the manual tagging of \yr{6188} and \scaleview’s runtime data, which
shows the buffer growing linearly with the number of decommissioning
nodes.

\section{Evaluation of \scalelens}
\label{sec-evaluation}

This evaluation aims to answer the following questions:
\begin{rqs}
\item How effective is \scalelens in identifying previously known 
real-world scalability faults, and how does it compare to the baseline?
\item Can \scalelens detect previously unknown scalability faults in the latest stable versions of distributed systems?

\end{rqs}

\subsection{Experimental Setup} \label{sec-evaluation-setup}

\subsubsection{Benchmark} As far as we know, there is no existing
benchmark that targets scalability faults in distributed systems.
Hence, we build our own based on the scalability faults we collected
for our empirical study (\cref{sec-empirical-methodology}). From the
\numbugs scalability faults, we focus on those reported for \sysca,
\syshd, and \sysig, the three systems with the most scalability faults
(\calculate{67+105+45} combined, as shown in \cref{tab-dim-systems})
and the most up-to-date documentation. 
Then, we \textbf{(1)} determine if the system's versions 
related to the faults are deployable and runnable, 
and \textbf{(2)} select the faults that are within the scope of \scalelens, \ie \textit{compute} and \textit{unbound} faults. The steps above 
result in \benchtotal real-world scalability issues, as listed in 
\cref{tab-eval-full}.

\subsubsection{Baseline}
The most closely related work to \scalelens is
\scalecheck~\cite{DBLP:conf/fast/StuardoLSKLCLG19}, an approach
designed to discover scalability faults. \scalecheck has two
components: \sfind and \stest. Among the two, \sfind is responsible
for detection, exposing \emph{scale-dependent loops} through program
analysis. Such loops are a narrower form of \dcfs. Since both \sfind
and \scaleview identify code fragments with scale-correlated iteration
counts from source code and workloads, we use \sfind as the baseline
for \scaleview. Technical details of \scalecheck and a head-to-head
comparison with \scalelens are provided in \cref{sec-related-work}.

\subsubsection{Workloads} We implemented $8$ workloads 
for \sysca, $3$ for \syshd, and $2$ for \sysig.
In those, we scaled a total of $8$ dimensions (\textit{nodes}, \textit{tokens}, 
\textit{tables}, \textit{rows} for \sysca, \textit{datanodes}, 
\textit{datablocks}, \textit{files} for \syshd, and \textit{rows} for \sysig)
and exercised a total of $6$ APIs (\textit{snapshot}, 
\textit{compaction}, \textit{repair} for \sysca, 
\textit{snapshot}, \textit{snapshotDiff} for \syshd, 
and \textit{query} for \sysig). The dimensions and APIs are selected based on their popularity, while \scalelens users can select any dimension or API of interest.
Each workload comprises $80-100$ lines of code in shell scripts and is
reutilized for every version we test. The same workloads are given to
\sfind and \scaleview.

\subsubsection{Environment} All experiments are conducted on a single
machine with 40-core \texttt{Xeon Platinum 8380} CPU and 256 GB of RAM. All
experiments, including the benchmark, are packaged into Docker images and are
publicly available to ensure the reproducibility of the results.

\subsection{RQ1: Effectiveness and Comparison with Baseline}

\label{sec-eval-effectiveness}

The goal of RQ1 is to evaluate the effectiveness of \scalelens in
detecting known scalability faults, and to provide a comparison with
the baseline \sfind.

We evaluate \scalelens on our benchmark of \benchtotal real-world
scalability faults drawn from our empirical study. For each fault, we
examine whether \scaleview detects the corresponding \dcf, and whether
\scalepick identifies the correct anti-pattern associated with that
\dcf. As summarized in \cref{tab-eval-full}, \scaleview detects the
correct \dcfs for 38 out of 55 faults, including both explicit and
implicit iterations. For each detected \dcf, \scaleview also correctly
identifies the associated system dimensions (not shown in
\cref{tab-eval-full} due to space constraints). We found that all $17$
cases in which \scaleview fails to detect \dcfs are related to
multi-threading. For example, \hd{13768} is a scalability fault where
all threads are forced to wait for the slowest to complete, a scenario
that requires deeper concurrency analysis. \scaleview is unable to
detect such instances whose root cause relates to the multi-threading
strategy.

On top of the \dcfs detected by \scaleview, \scalepick successfully
identifies anti-patterns for 36 out of 38 cases, demonstrating high
effectiveness. In the remaining 2 cases, the \dcfs are enclosed in
inner classes implementing the \texttt{java.lang.Runnable} interface.
Such classes are usually designed for multi-threading, being submitted
to thread pools for executors, and are not handled by the call graph
analysis in \scalepick. We leave as future work the detection of
scalability faults related to multi-threading.

We also compare \scalelens and \sfind in their capability to detect
\dcfs (\sfind does not detect anti-patterns). As shown in
\cref{tab-eval-full}, \sfind identified \dcfs for only 9 out of the
\benchtotal faults, all of which involve explicit iterations. Compared
to \scalelens, \sfind detected 29 fewer \dcfs. Even restricting our
comparison to explicit iterations, \scalelens still outperforms \sfind
by detecting 18 more \dcfs.

There are two reasons why \scaleview, the dynamic analysis component
of \scalelens, outperforms \sfind in \dcf detection. First, \sfind is
designed to monitor the state of the system by performing heap
measurement, which makes it unable to detect scale-dependent data
structures that are not persistent when dimensions scale. In
contrast, \scaleview directly instruments the source code to obtain
\dcfs. Second, by design, \sfind only considers explicit application
iterations, whereas \scaleview takes into account both explicit
application or library iterations and implicit iterations.

In addition to detecting more faults, \scalelens provides richer
diagnostic information. While \sfind only reports the presence of a
scale-dependent loop, \scalelens reveals (1) the system dimension
associated with the \dcf, and (2) the anti-pattern underlying the
\dcf. These insights are crucial for developers to understand the root
cause of scalability faults.

\vspace{0.5em}
\begin{tcolorbox}[boxsep=2pt,left=2pt,right=2pt,top=1pt,bottom=1pt]
\textbf{Response to RQ1}: \scalelens detects 29 more (4.2\texttimes)
\dcfs than the baseline, and fully identifies the \dcfs and
anti-patterns for 36 out of \benchtotal previously known scalability
faults. \scalelens also reveals richer diagnostic information, such as
the correlated system dimensions.
\end{tcolorbox}

\subsection{RQ2: Finding Unknown Scalability Faults}
\label{sec-eval-implications}

In RQ1, we discussed the effectiveness of \scalelens in detecting
previously known scalability faults included in our empirical study.
The goal of RQ2 is to evaluate the effectiveness of \scalelens in detecting
previously unknown scalability faults not present in our study. For that, we focus this evaluation on the latest stable versions of \sysca (4.1.0), \syshd
(3.4.0), and \sysig (2.16.0).

\subsubsection{Ablation Analysis \& Manual Inspection}
\scalelens combines \scaleview and \scalepick. To assess the
contribution of each component, we analyzed the results from running
\scalelens on the latest stable versions of \sysca, \syshd, and
\sysig, as detailed in \cref{tab-ablation}.

Without the analysis of \scalelens, all loops and methods in the
codebase are potential \dcfs, leading to a large number of code
fragments: 68,966 in \sysca, 74,737 in \syshd, and 190,059 in
\sysig.\footnote{Loops represent explicit iteration while a method
called multiple times would capture implicit iteration.} Simply
applying \scalepick to detect static code anti-patterns results in
7,823 to 16,982 code fragments, still making manual inspection
impractical.

\begin{table}[t]
    \centering
    \scriptsize
    \caption{Ablation Analysis of \scalelens. }
    \begin{tabular}{llll}
        \toprule
        & \sysca & \syshd & \sysig \\
        \midrule
        \# Fragments                 &   68,966                      & 74,737                   & 190,059                   \\
        \# Fragments + Anti-Patterns & 7,823                         & 16,982                   & 12,467                    \\
        \midrule
        \# \dcfs                      & 251 (\ratio{251}{68966})                           & 106 (\ratio{106}{74737})                      & 342 (\ratio{342}{190059})                        \\
        \# \dcfs + Anti-Patterns      & 129 (\ratio{129}{68966})                           & 68 (\ratio{68}{74737})                       & 137 (\ratio{137}{190059})     \\
        \bottomrule
\end{tabular}
    \label{tab-ablation}
\end{table}

\scaleview significantly reduces the number of code fragments of
interest from 68,966 to 251 for Cassandra, representing
\ratio{251}{68966} of the whole system. Out of the 251 \dcfs
pinpointed, \scalepick further identifies 129 that are associated with
anti-patterns. Similarly, \scaleview reduces the number of code
fragments from 74,737 to 106 for \syshd, and from 190,059 to 342 for
\sysig, with 68 and 137 \dcfs containing anti-patterns, respectively.
To confirm the correctness of the \dcfs, we manually inspected all
\dcfs by plotting the iteration count to verify the increase of
computational overhead along with the increase of the scale of the
flagged dimension. For the \dcfs with anti-patterns, we located the
anti-patterns in the source code and confirmed their existence.

\subsubsection{Scalability Faults in the Wild}

\scalelens identifies 129, 68, and 137 \dcfs associated with anti-patterns in
the latest versions of \sysca, \syshd, and \sysig, respectively. From them, we
found that 28 \dcfs from \sysca and 7 from \syshd are related to
(de)serialization operations. Notably, all of them are (1) labeled as having
linear growth to the dimension, and (2) identified with the
\textit{compute-cross} anti-pattern, which is correct as such operations are
known to be \io-intensive due to their purposes.  Such (de)serialization
operations are often considered as necessary costs in distributed systems and
are often unavoidable without major redesigns.  As such, we exclude them from
reporting. For the remaining \dcfs, we manually confirmed that all of them
exhibit problematic behavior and should be reported to developers.

To contribute to the effort of
reporting previously unknown scalability faults, we have reported 27 \dcfs with
anti-patterns to the developers of \sysca and \syshd. So far, the developers
have confirmed problematic behavior for 5 of them and are investigating 4
others. The rest are pending review and none have been marked as invalid.  We
are in the process of reporting the rest of the scalability faults.

\vspace{0.5em}
\begin{tcolorbox}[boxsep=2pt,left=2pt,right=2pt,top=1pt,bottom=1pt]
  \textbf{Response to RQ2}: By combining dynamic and static analyses,
  \scalelens enables efficient detection of scalability
  faults. \scalelens has detected previously unknown faults in the
  latest stable versions of \sysca, \syshd, and \sysig, resulting so
  far in 27 reports to developers, with 5 confirmed and 4 under
  investigation.
\end{tcolorbox}
\section{Threats to Validity}
\label{sec-threats-to-validity}

\subsection{Construct Validity}
\label{sec-threats-construct}

The central constructs of our work are \emph{scalability fault},
\emph{scalable system dimension}, \emph{DCF}, and \emph{anti-pattern}.
We defined each at first use and anchored our definitions in prior
literature where possible: scalability fault follows
\citet{DBLP:conf/hotos/Leesatapornwongsa17}, and scalable system
dimensions follow the scalability characterization framework of
\citet{DBLP:conf/icse/DubocRW06, DBLP:conf/sigsoft/DubocRW07}. Alternative operational definitions are
possible. To guard against inconsistency, we applied the resulting
taxonomy uniformly across all \numbugs faults and released the full
catalog of labeled issues for independent audit.

We use iteration count as the observable proxy for scale-dependent
execution cost. However, some scale-sensitive costs do not manifest
as loops, such as a recursive descent whose depth correlates with a
dimension, or a chain of conditional branches whose path space grows
with cluster size. To reduce the risk of missing such cases, \scaleview tracks not only
explicit iteration but also implicit iteration, where the number of
method invocations across homogeneous components grows with the
system's scale (\cref{sec-approach-instrumentation}).

Last, a construct-validity concern is the fidelity of the manually
authored scaling workloads. A workload is intended to drive one or
more dimensions to a large value while exercising representative APIs;
a badly designed workload may fail to do so. We mitigate this by
requiring workloads to report their current scale state through a
\texttt{set-scale-state} API provided by \scaleview (\cref{sec-approach-workload}), which gives \scaleview a
runtime signal for whether the intended dimension is actually being
scaled, and by releasing the full set of workloads used in our
evaluation for independent inspection.

\subsection{Internal Validity}
\label{sec-threats-internal}

Our empirical findings rest on the manual selection and categorization
of \numbugs scalability faults from over \totalbugs issue reports.
This manual process can introduce bias. We mitigate this with a
multi-stage adjudication protocol: every issue is independently
reviewed and tagged by two authors, a third author reviews every
disagreement or uncertainty, and unresolved conflicts are settled in
group discussion. We further mitigate the threat by publishing the
full set of tagged issues and the tagging guidelines.

We also considered keyword-based selection as an alternative to this
manual process. A broad keyword search over the \totalbugs issues,
using terms such as ``scale'', ``scalability'', ``cluster'', ``size'',
``degradation'', ``contention'', ``out of memory'', ``performance'',
``regression'', and ``impact'', returned $10,046$ candidates, of
which manual inspection revealed a 99\% false-positive rate. Backward
validation against the manually selected \numbugs faults showed that
78\% of them contain none of those keywords. Automating selection by
keyword would therefore have both over-reported and under-reported
scalability faults, and the manual process was the only reliable
option available.

A further internal-validity concern is the inherent nondeterminism of
dynamic analysis: the JIT compiler, garbage collector, and operating
system scheduler introduce run-to-run variation in \scaleview's
recorded iteration counts. We mitigate this by filtering growth
trends across multiple scale states rather than relying on
single-point measurements (\cref{sec-approach-filtering}), so that
the correlation between iteration count and scale state is robust to
local noise.

\subsection{External Validity}
\label{sec-threats-external}

The \numrepos systems in our empirical study
(\cref{sec-empirical-fault-patterns}) are all Java-based and span
databases, file systems, compute frameworks, streaming platforms, and
coordination services. This covers the most common categories of
open-source distributed systems, but our findings may not transfer to
non-JVM distributed systems (\eg C++, Go, or Rust) or to application
domains with substantially different scaling profiles, such as
HPC-style tightly coupled compute.

\scalelens has two components with different generalization
properties. \scaleview instruments code at the JVM bytecode level and
is therefore applicable to any Java-based system without source
modification. \scalepick, in contrast, depends on static call-graph
analysis, which is accurate for code with conventional control flow
(explicit method calls, inheritance, explicit locks) but is less
precise on event-driven, callback-heavy, reflective, or
executor-dispatched code, where the call graph is incomplete or
statically unresolvable. In our evaluation, for instance, two
anti-pattern false negatives trace to \texttt{Runnable}-based
executor code, which our analysis does not follow into the thread
pool. In practical terms, \scalelens is most effective on Java batch,
storage, and database systems such as \sysca, \syshd, and \sysig,
where control flow and synchronization relationships are largely
recoverable from the source. Extending \scalelens to event-driven
architectures would likely require hybrid static-dynamic call-graph
recovery, which we leave as future work.

\scalelens's detection performance additionally depends on the
coverage of the supplied scaling workloads. A DCF is detected only if
the relevant dimension is exercised, so workloads that omit a
dimension will not surface DCFs that depend on it. This is a
generalization threat rather than a tool bug. Users of \scalelens can
mitigate it by authoring workloads that jointly scale the dimensions
they care about, and by reusing workloads across versions; in our
evaluation the same workloads were reused across 7 versions of
\sysca, 5 of \syshd, and 3 of \sysig at no additional cost.

For the tool evaluation, our benchmark comprises \benchtotal
scalability faults drawn from \sysca, \syshd, and \sysig, which
together account for $217$ (\ratio{217}{444}) of the faults in our
study. This gives substantial coverage but is not exhaustive: the
remaining seven studied systems and fault categories outside
\textit{compute} and \textit{unbound} are not represented in the
benchmark. We additionally explored using the 10-bug benchmark from
\scalecheck~\cite{DBLP:conf/fast/StuardoLSKLCLG19}, but those
systems (\sysca 1.1.x and 1.2.x, \syshd 2.0.0) were released before
April 2013 and are no longer buildable due to outdated dependencies;
we therefore report numbers only against our own benchmark.

Finally, the taxonomy itself carries an external-validity concern. The
4 root-cause categories and the 11 anti-patterns emerged from \numbugs
faults in \numrepos systems, so additional systems or application
domains, particularly outside the Apache Java ecosystem, may surface
different categories or anti-patterns.

\subsection{Conclusion Validity}
\label{sec-threats-conclusion}

\scaleview finds DCFs as system facts: a DCF is by construction an
iterative code fragment whose iteration count correlates with a scaled
dimension, so \scaleview produces no false positives for DCF
identification. \scaleview may, however, miss DCFs whose dimension is
not exercised by the supplied workload, a source of false negatives
already discussed above.

\scalepick flags a \dcf with an anti-pattern only when its slice
contains a checker-specific sink (\cref{section-static-checker}). We
manually verified every flagged anti-pattern on our benchmark and
observed no false positives. The two anti-pattern false negatives on
known faults both trace to \texttt{Runnable}-based executor code,
tying back to the static-analysis limitation noted under External
Validity.

Our headline effectiveness numbers, 36 fully detected and 2 partially
detected out of \benchtotal known scalability faults and
4.2\texttimes\xspace more DCFs than \sfind, are reported against this
benchmark. They support our claims about \scalelens's detection
capability on the studied fault categories and systems, but should
not be read as universal guarantees across arbitrary distributed
systems. The \calculate{129+68+137} DCFs with anti-patterns reported
on the latest stable versions of \sysca, \syshd, and \sysig were
individually verified by manual inspection, so the positive count is
reliable, but the proportion that developers will ultimately confirm
as problematic is still evolving as reports progress through upstream
review.

\section{Related Work}
\label{sec-related-work}

\scalecheck~\cite{DBLP:conf/fast/StuardoLSKLCLG19} discovers
scalability bugs in large-scale distributed systems via two
components: \sfind, a program analysis tool that identifies
scale-dependent loops (those whose number of iterations grow alongside
system-scale data structures), and \stest, a set of colocation
techniques that emulate a real-scale cluster on a single machine.
\scalelens differs from \scalecheck in three key ways. First,
\scaleview generalizes scale-dependent loops into \dcfs by capturing
implicit iteration, even when no explicit loop exists in the source.
Second, while \sfind tracks heap-allocated collections to find loops,
\scaleview instruments bytecode directly; this allows it to detect
\dcfs involving transient data structures that a heap trace would
miss. Finally, whereas \scalecheck relies on \stest to exercise the
identified loops through cluster emulation, \scalelens provides
\scalepick, a set of static checkers to identify anti-patterns known
to incur in scalability issues when combined with \dcfs. These
approaches are complementary: \scalepick offers immediate root-cause
analysis, while \stest reproduces problematic symptoms at runtime.

Other tools
also follow this symptom-oriented emulation approach \cite{
  DBLP:conf/nsdi/0009KSAD14, DBLP:conf/nsdi/GuptaVV08,
  DBLP:conf/opodis/MachadoMNC019, DBLP:conf/icfc/ZengCS19}. For example,
\textsc{Exalt} \cite{DBLP:conf/nsdi/0009KSAD14} aims to emulate
hundreds of \syshd nodes in a single machine to observe I/O-related
increases in processing time, while \textsc{DieCast}~\cite{DBLP:conf/nsdi/GuptaVV08} aims to emulate large networks using a
single machine and observe how networking speed affects processing
time.  In contrast, \scalelens does not require explicit symptoms nor
emulation techniques as it detects unknown scalability faults based on
a combination of \dcfs and anti-patterns.

Other works utilize extrapolation
\cite{DBLP:conf/mascots/ShiGW18,
  DBLP:conf/ipps/ZhouZSS20,
  DBLP:journals/cacm/LagunaASGLSBKZC15,
  DBLP:conf/hpdc/ZhouKB11,
  DBLP:journals/tpds/JinWTGZHLLZ25}
to project system behavior (\eg execution time)
at scale using smaller-scale measurements.
\textsc{PatternMiner} \cite{DBLP:conf/mascots/ShiGW18} identifies
scalability bottlenecks in centralized distributed
systems (\eg \syshd) by detecting repeating behavior and noting that
such behavior repeats at larger scales.
\textsc{Vrisha}~\cite{DBLP:conf/hpdc/ZhouKB11},
\textsc{ScalAna}~\cite{DBLP:journals/tpds/JinWTGZHLLZ25},
\textsc{AutomaDeD}~\cite{DBLP:journals/cacm/LagunaASGLSBKZC15} and
\cite{DBLP:conf/ipps/ZhouZSS20} model the scalability of HPC
applications combining smaller-scale behavior observations with
techniques such as canonical correlation analysis, performance graphs,
stack trace analysis and machine learning, respectively. \scalelens
differs in that it requires no specification of symptoms/behaviors, is
architecture-agnostic, and focuses on distributed systems.

Finally, other works present techniques for detecting memory bloat
\cite{ DBLP:conf/wcre/JezekL17, DBLP:journals/software/MitchellSS10,
  DBLP:conf/iwmm/BuBXC13, DBLP:conf/cgo/Li0CJ023,
  DBLP:journals/tosem/NguyenWBFX18}, resource leaks \cite{
  DBLP:conf/sigsoft/KelloggSSE21, DBLP:journals/corr/abs-2311-04448,
  DBLP:conf/pldi/XuBQR11, DBLP:journals/tosem/XuR13,
  DBLP:conf/kbse/ShahoorKYK23}, thread contention
\cite{DBLP:journals/access/ZhangLLL19, DBLP:conf/icmla/LiscanoARASC23,
  DBLP:conf/osdi/ZhouGMW18, DBLP:conf/osdi/AhnHKJ24, DBLP:conf/issta/YuP16, DBLP:journals/tpds/LiGHLZLH24, DBLP:conf/eurosys/AlamLZM17}, chatty I/O \cite{chatyiomicrosoft.web,
  DBLP:conf/icse/YangYWLC19,  DBLP:conf/icsa/PinciroliAT23,
  DBLP:conf/icsa/AvritzerJTRCMO25}, and performance anti-patterns
  \cite{DBLP:conf/wosp/SmithW00,
    DBLP:conf/cmg/SmithW02b,
    DBLP:conf/icse/ChenSJHNF14,
    DBLP:conf/kbse/ChenJML19,
    DBLP:conf/icsm/ShaoQYYJXW20,
    DBLP:journals/tse/TrubianiPBF23, DBLP:conf/msr/ZhaoGZHTC24,
  DBLP:conf/eurosys/LiCLLZGGLL18,
  DBLP:conf/icse/SongL17, DBLP:conf/icse/NistorSML13,
  DBLP:conf/icse/NistorCRL15, petriu1997pattern}. We consider all these works orthogonal to \scalelens, as they are limited to design flaws regardless of system scale. \scalelens, however, focuses on scalability faults unique to large-scale distributed systems, identifying anti-patterns whose negative impact remains latent at small scales but manifests at large scales due to growth in one or more system dimensions.

\section{Conclusion}
\label{sec-conclusion}

To better understand scalability faults, we conducted a systematic
study of \numbugs scalability faults from 10 distributed systems. Our
study uncovered the notion of \dcfs, and various anti-patterns that
result in scalability faults when combined with \dcfs. Based on this,
we developed \scalelens, a novel approach that combines dynamic and
static analyses to detect scalability faults. Our evaluation showed
that \scalelens effectively finds previously known scalability faults
(36 out of 55), outperforms the baseline (4.2\texttimes\xspace more
\dcfs), and correctly detects \calculate{129+68+137} \dcfs associated with anti-patterns
in the latest stable versions of \sysca, \syshd, and
\sysig. The source code of \scalelens is publicly available at
\url{https://github.com/ucd-plse/scalelens}. The empirical study
data, evaluation data, and reproduction instructions are at
\url{https://github.com/ucd-plse/scalability}.

\section*{Acknowledgment}
This work is supported in part by NSF grants CCF-2119348 and CCF-2119184. Results presented in this paper were obtained using the Chameleon testbed~\cite{DBLP:conf/usenix/KeaheyAZRRSCCGH20} supported by the National Science Foundation.

\balance

\bibliographystyle{abbrvnat-no-isbn}
{\small \bibliography{references,related-work,reference-weblinks}}

\end{document}